\documentclass[english,aps,preprint]{revtex4}
\usepackage[T1]{fontenc}
\usepackage[latin9]{inputenc}
\usepackage{graphicx}

\usepackage{babel}

\begin{document}

\title{Pair Production of Fourth Family Charged Sleptons at $e^{+}e^{-}$
Colliders}

\author{V. Ar\i{}}

\email{vari@science.ankara.edu.tr}

\affiliation{Ankara University, Faculty of Sciences, Department of Physics, 06100,
Tandogan, Ankara, Turkey}

\author{\.{I}. T. \c{C}ak\i{}r}

\email{tcakir@mail.cern.ch}

\affiliation{Department of Physics, CERN, Geneva, Switzerland}

\author{O. \c{C}ak\i{}r}

\email{ocakir@science.ankara.edu.tr}

\affiliation{Ankara University, Faculty of Sciences, Department of Physics, 06100,
Tandogan, Ankara, Turkey}

\author{S. Sultansoy}

\email{ssultansoy@etu.edu.tr}

\affiliation{Physics Division, TOBB University of Economics and Technology, Ankara,
Turkey }

\affiliation{Institute of Physics, Academy of Sciences, Baku, Azerbaijan}
\begin{abstract}
We study the pair production of $\tilde{l}_{4}$, which is the supersymmetric
partner of the fourth family charged lepton, at the $e^{+}e^{-}$
colliders. It is shown that the investigation of this process at ILC/CLIC
will give opportunity to differentiate the MSSM with three and four
families.
\end{abstract}
\maketitle

Recently, the fourth Standard Model (SM4) family attracts an increasing
interest of high energy physics (HEP) community \cite{3SM_1,3SM_2}
(see also \cite{Holdom10} and ref's therein). In some sense, we deal
with SM4 {}``renaissance''. Actually, the existence of the fourth
SM family follows from the SM basics \cite{Fritzsch92,Datta93,Celikel95}
(see also review \cite{Sultansoy09} and ref's therein). Moreover,
former objection against SM4 based on electroweak precision data is
disappeared \cite{Kribs07}: new studies show that SM3 and SM4 has
the same status (see \cite{Cobanoglu10} and ref's therein). The current
limits on the mass of the fourth SM family leptons are \cite{Nakamura10}:
$m_{l_{4}}>100$ GeV, $m_{\nu_{4}}>90$ ($80$) GeV for Dirac (Majorana)
neutrinos. Recently, the Collider Detector at Fermilab (CDF) has already
constrained the masses of fourth family quarks: $m_{u_{4}}>335$ GeV
at 95\% CL. \cite{Conway}, $m_{d_{4}}>338$ GeV at 95\% CL. \cite{Aaltonen}.
On the other hand, the partial wave unitarity leads to an upper bound
$700$ GeV for fourth SM family fermion masses \cite{Chanowitz09}. 

The SM, while well describing almost all experimental HEP data, has
not answered a number of fundamental questions. For this reason, different
approaches beyond the SM are developed. Among them, the supersymmetry
(SUSY) plays an important role. Below, we denote this minimal supersymmetric
model as MSSM4 (MSSM3) in the case of four (three) SM families. Within
the MSSM3, for example the low mass eigenstate stop $\tilde{t}_{1}$
may be lighter than the top quark due to the large Yukawa coupling
\cite{Ellis83}. By analogy, within the MSSM4 the fourth family sfermions
could be lighter than coresponding fermions. Because of the larger
masses of the fourth family fermions (expected around $500$ GeV)
the fourth family sfermions quite may be the lightest ones among the
squarks and sleptons. It is difficult to differentiate MSSM3 and MSSM4
at hadron colliders, because the light superpartners of the third
and fourth family quarks has almost the same decay chains. 

In this work, we consider pair production of supersymmetric fourth
family charged sleptons, which is expected to be the lightest one
among the charged sleptons, at future linear colliders, namely ILC
with $\sqrt{s}=1$ TeV {[}ILC] and CLIC with $\sqrt{s}=3$ TeV {[}CLIC].
We present the results for pair production of the supersymmetric partner
of the fourth family charged slepton $\tilde{l}_{4}$ (also called
stau-prime) decaying to fourth family neutrino $\nu_{4}$ and a chargino
$\tilde{\chi}_{1}^{-}$ with a subsequent $\tilde{\chi}_{1}^{-}$
decay into a neutralino $\tilde{\chi}_{1}^{0}$ and $W^{-}$ boson.

The inclusion of the fourth SM family into MSSM is straightforward
\cite{Carena96}. Concerning the mass matrix of the fourth family
sfermions, let us consider the charged sleptons case. In the $(\tilde{l}_{4L},\tilde{l}_{4R})$
basis, the mass matrix is given by\begin{equation}
M_{\tilde{l}_{4}}^{2}=\left(\begin{array}{cc}
m_{\tilde{l}_{4L}}^{2} & a_{l_{4}}m_{l_{4}}\\
a_{l_{4}}m_{l_{4}} & m_{\tilde{l}_{4R}}^{2}\end{array}\right)\label{eq:1}\end{equation}
where $m_{\tilde{l}_{4L}}^{2}=M_{\tilde{L}_{4}}^{2}+m_{l_{4}}^{2}-m_{Z}^{2}$$\cos2\beta(\frac{1}{2}-\sin^{2}\theta_{W})$,
$m_{\tilde{l}_{4R}}^{2}=M_{\tilde{E}{}_{4}}^{2}+m_{l_{4}}^{2}-m_{Z}^{2}$$\cos2\beta\sin^{2}\theta_{W}$
and $a_{l_{4}}=A_{l_{4}}-\mu\tan\beta$, $A_{l_{4}}$ is trilinear
Higgs-fourth family charged lepton parameter (the notation of \cite{Bartl05}
is used, extension to MSSM4 is straightforward). 

The mass eigenstates $\tilde{l}{}_{4l}$ and $\tilde{l}{}_{4h}$ are
related to $\tilde{l}{}_{4L}$ and $\tilde{l}{}_{4R}$ by

\begin{equation}
\left(\begin{array}{c}
\tilde{l}{}_{4l}\\
\tilde{l}_{4h}\end{array}\right)=\left(\begin{array}{cc}
\cos\theta_{\tilde{l_{4}}} & \sin\theta_{\tilde{l_{4}}}\\
-\sin\theta_{\tilde{l_{4}}} & \cos\theta_{\tilde{l_{4}}}\end{array}\right)\left(\begin{array}{c}
\tilde{l}_{4L}\\
\tilde{l}_{4R}\end{array}\right)\label{eq:2}\end{equation}
with the eigenvalues 

\begin{equation}
m_{\tilde{l}_{4(l,h)}}^{2}=\frac{1}{2}(m_{\tilde{l}{}_{4L}}^{2}+m_{\tilde{l}{}_{4R}}^{2})\mp\frac{1}{2}\sqrt{(m_{\tilde{l}{}_{4L}}^{2}-m_{\tilde{l}{}_{4R}}^{2})^{2}+4a_{l_{4}}^{2}m_{l_{4}}^{2}}\label{eq:3}\end{equation}
 and the mixing angle $\theta_{\tilde{l}_{4}}$ is given by

\begin{equation}
\cos\theta_{\tilde{l}_{4}}=\frac{-a_{l_{4}}m_{l_{4}}}{\sqrt{(m_{\tilde{l}{}_{4L}}^{2}-m_{\tilde{l}{}_{4l}}^{2})^{2}+a_{l_{4}}^{2}m_{l_{4}}^{2}}}\label{eq:4}\end{equation}
As seen from Eq. (\ref{eq:3}), $\tilde{l}{}_{4l}$ is expected to
be the lightest charged slepton because of large value of $m_{l_{4}}$. 

The cross section for the process $e^{+}e^{-}\to\tilde{l}_{4l}^{+}\tilde{l}_{4l}^{-}$
is:

\begin{eqnarray}
\sigma(e^{+}e^{-} & \rightarrow & \tilde{l}_{4l}^{+}\tilde{l}_{4l}^{-})=\frac{\pi\alpha^{2}}{3s^{4}}(s^{2}-4sm_{\tilde{l}_{4l}}^{2})^{3/2}\nonumber \\
 & \times & \left(1+\frac{a_{ll}v_{e}s}{8\sin^{2}\theta_{W}\cos^{2}\theta_{W}}Re[D(Z)]+\frac{a_{ll}^{2}(v_{e}^{2}+a_{e}^{2})s^{2}}{256\sin^{4}\theta_{W}\cos^{4}\theta_{W}}|D(Z)|^{2}\right)\end{eqnarray}
where $D(Z)=1/[(s-m_{Z}^{2}+i\Gamma_{Z}m_{Z}];$ $v_{e}=-1+4\sin^{2}\theta_{W}$,
$a_{e}=-1$ and $a_{ll}=2(-\cos^{2}\theta_{\tilde{l}_{4}}+2\sin^{2}\theta_{W})$. 

There are two main decay modes of $\tilde{l}_{4l}$, namely $\tilde{l}_{4l}^{-}\rightarrow\nu_{4}\tilde{\chi}_{1}^{-}$
and $\tilde{l}_{4l}^{-}\rightarrow\tilde{\nu}_{4}W^{-}$ . Here, we
are interested in the first decay channel since it allows to differentiate
the signal of $\tilde{l}_{4l}$ from lighter stau $\tilde{\tau}_{1}$.
The chargino decays dominantly through $\tilde{\chi}_{1}^{-}\rightarrow\tilde{\chi}_{1}^{0}W^{-}$
where we assume the $\tilde{\chi}_{1}^{0}$ to be the LSP. The $\nu_{4}$
decays are determined by the nature of the fourth family neutrino.
In the Dirac case we deal with $\nu_{4}\rightarrow l^{-}W^{+},$ while
in the Majorana case we also have $\nu_{4}\rightarrow l^{+}W^{-}$
in addition to $\nu_{4}\rightarrow l^{-}W^{+},$ with the branchings
$BR(\nu_{4}\rightarrow l^{+}W^{-})=BR(\nu_{4}\rightarrow l^{-}W^{+}).$ 

\begin{figure}
\includegraphics{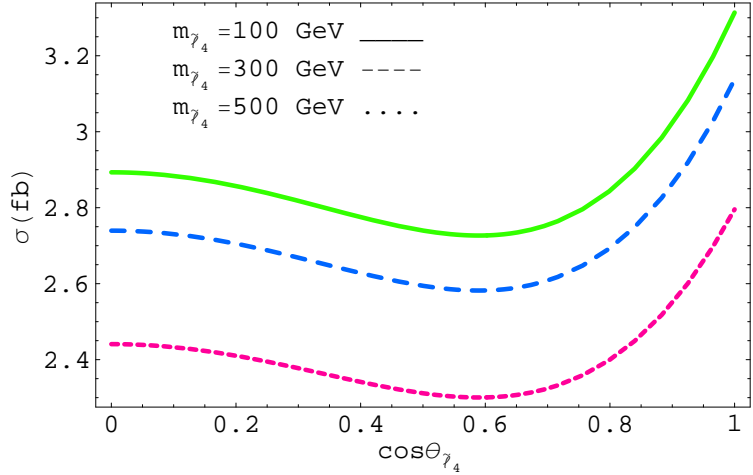}

\caption{The cross section depending on the mixing parameter for different
masses $m{}_{\tilde{l}_{4}}$ at $\sqrt{s}=3$ TeV.\label{fig:fig1}}

\end{figure}

\begin{figure}
\includegraphics{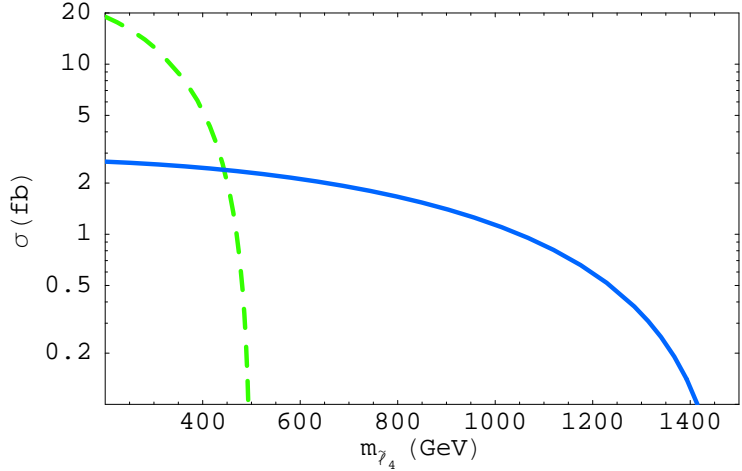}

\caption{The cross sections depending on the mass of the fourth family charged
lepton at 1 TeV (dashed) and 3 TeV (solid) center of mass energies.
\label{fig:fig2}}

\end{figure}

In Fig. \ref{fig:fig1}, we present the cross section depending on
the mixing parameter for different masses at $\sqrt{s}=3$ TeV. It
is seen that the cross section has a minimum at $\cos\theta_{\tilde{l}_{4}}=0.6$
and maximum at $\cos\theta_{\tilde{l}_{4}}=1$. Hereafter, we use
the conservative value of $\cos\theta_{\tilde{l}_{4}}=0.6$ for numerical
calculations. The cross sections depending on the mass $m_{\tilde{l}_{4}}$
for two different values of center of mass energies are presented
in Fig. \ref{fig:fig2}. Taking the fourth family charged slepton
mass $m_{\tilde{l}_{4}}=300$ GeV at ILC with $\sqrt{s}=1$ TeV, we
obtain the cross section $\sigma=12.6$ fb. At CLIC with 3 TeV center
of mass energy we obtain the cross section $\sigma=2.3$ fb for $m_{\tilde{l}_{4}}=500$
GeV. As seen from Fig. \ref{fig:fig2}, ILC is advantageous up to
the mass $m_{\tilde{l}_{4}}\simeq450$ GeV.

We propose the following decay chains for the signal if the fourth
family neutrino has the Dirac nature:

\begin{equation}
\tilde{l}_{4}^{-}\to\nu_{4}(\to\mu^{-}W^{+}(\to hadrons))+\tilde{\chi}_{1}^{-}(\to\tilde{\chi}_{1}^{0}W^{-}(\to\mu^{-}\nu))\label{eq:6a}\end{equation}

\begin{equation}
\tilde{l}_{4}^{+}\to\bar{\nu}_{4}(\to\mu^{+}W^{-}(\to\mu^{-}\nu))+\tilde{\chi}_{1}^{+}(\to\tilde{\chi}_{1}^{0}W^{+}(\to hadrons))\label{eq:6b}\end{equation}
This signal will be seen in detector as $3\mu^{-}+1\mu^{+}+4j+E_{T}^{miss}$.
The decays of the $\nu_{4}$ are governed by the leptonic $4\times4$
mixing matrix. For numerical calculations we use the parametrization
given in \cite{Ciftci05}, which is compatible with the experimental
data on the masses and mixings in leptonic sector. In this case, $BR(\nu_{4}\to\mu^{-}W^{+})\simeq0.68$
and $BR(\nu_{4}\to\tau^{-}W^{+})\simeq0.32$, which is the reason
for choosing the muon channel for $\nu_{4}$ decays. Keeping in mind
that $BR(W^{-}\to l^{-}\nu)=0.11$ and $BR(W^{-}\to hadrons)=0.68$
and assuming that $BR(\tilde{l}_{4l}\rightarrow\nu_{4}\tilde{\chi}_{1}^{-})=0.5,$
we obtain resulting branching for the considered signal as $6.5\times10^{-4}$.
This corresponds to 8 signal events for $m_{\tilde{l}_{4}}=300$ GeV
at ILC with $\sqrt{s}=1$ TeV and $L_{int}=1$ ab$^{-1}$. It should
be noted that this signal is almost backgroundless. Furthermore, the
number of signal events is doubled if we consider the charge conjugated
process. Additional factor 4 comes from counting the $W$- decays
into electron channel. 

The CLIC with the center of mass energy $\sqrt{s}=$3 TeV and integrated
luminosity $L_{int}=1$ ab$^{-1}$ will give opportunity to investigate
fourth family charged slepton with masses up to $1$ TeV. In the Dirac
$\nu_{4}$ case, for the pair production of $\tilde{l}_{4l}$ with
500 GeV mass, we obtain 2 signal events in the channel $3\mu^{-}+1\mu^{+}+4j+E_{T}^{miss}$
and 4 events when charge conjugated process is added, and 16 events
if the $W\to e\nu$ channels are also considered. 

Even more spectacular signature is expected if the $\nu_{4}$ has
the Majorana nature. In this case, $\nu_{4}$ will decay into both
$\mu^{-}W^{+}$ and $\mu^{+}W^{-}$ with the same branching ratio
equal to 0.34. If $\bar{\nu}_{4}$ in Eq. (\ref{eq:6b}) decays into
$\mu^{-}W^{+}(\rightarrow hadrons)$ channel, the signal will be seen
in dedector as $3\mu^{-}+6j+\mbox{\ensuremath{\mbox{\ensuremath{E_{T}^{miss}}}}}$.
This signal is free of background and has resulting branching equal
to $1.0\times10^{-3}.$ This corresponds to 12 signal events for $m_{\tilde{l}_{4}}=300$
GeV at ILC with $\sqrt{s}=1$ TeV and $L_{int}=1$ $\mbox{ab\ensuremath{^{-1}}. The number of events becomes 24 if charge conjugated process is addded. }$

Concerning the CLIC with $\sqrt{s}=3$ TeV and $L_{int}=1$ ab$^{-1}$
integrated luminosity we obtain 5 signal events for $m_{\tilde{l}_{4}}=500$
GeV when charge conjugated process is added, and 10 signal events
if the $W^{+}\to e^{+}\nu$ channel is counted as well.

In conclusion, it is quite possible that the first manifestation of
MSSM4 will come from the lepton colliders. This case should be seriously
analysed within the physics program of future lepton colliders.

\end{document}